\documentclass[showpacs,preprintnumbers,preprint,amsmath,amssymb]{revtex4-1}

\usepackage{graphicx}
\usepackage{dcolumn}
\usepackage{bm}
\usepackage[shortlabels]{enumitem}
\begin{document}

\preprint{LA-UR-13-20814}

\title{Application of vibration-transit theory to distinct dynamic response for a monatomic liquid}
\author{Duane C. Wallace}
\author{Eric D. Chisolm}
\author{Giulia De Lorenzi-Venneri}

\affiliation{Theoretical Division, Los Alamos National Laboratory, 
Los Alamos, New Mexico 87545}

\date{\today}

\begin{abstract}
We examine the distinct part of the density autocorrelation function $F^{d}(q,t)$, also called the intermediate scattering function, from the point of view of the vibration-transit (V-T) theory of monatomic liquid dynamics. A similar study has been reported for the self part, and we study the self and distinct parts separately because their damping processes are not simply related. We begin with the perfect vibrational system, which provides precise definitions of the liquid correlations, and provides the vibrational approximation $F^{d}_{vib}(q,t)$ at all  $q$ and  $t$. Two independent liquid correlations are defined, motional and structural, and these are decorrelated sequentially, with a crossover time $t_{c}(q)$. 
This is done by two independent decorrelation processes:
the first, vibrational dephasing, is naturally present in $F^{d}_{vib}(q,t)$ and operates to damp the motional correlation; the second, transit-induced decorrelation, is invoked to \emph{enhance} the damping of motional correlation, and then to damp the structural correlation. A microscopic model is made for the ``transit drift," the averaged transit motion that damps motional correlation on  $0 \leq t \leq t_{c}(q)$. Following the previously developed self-decorrelation theory, a microscopic model is also made for the ``transit random walk," which damps the structural correlation on  $t \geq t_{c}(q)$. The complete model incorporates a property common to both self and distinct decorrelation:  simple exponential decay following a delay period, where the delay is $t_{c}(q)$, the time required for the random walk to emerge from the drift.  Our final result is an accurate expression for $F^{d}(q,t)$ for all $q$ through the first peak in $S^{d}(q)$. (A modification will be required at $q$ where $S^{d}(q)$ converges to zero.) The theory is calibrated and tested using molecular dynamics (MD) calculations for liquid Na at 395~K; however, the theory itself does not depend on MD, and we consider other means for calibrating it.
\end{abstract}

\maketitle

\section{Introduction}

Our goal  in developing V-T theory is to apply the established techniques of many body physics to the mechanical problem of the motion of atoms in a monatomic liquid. The many body formulation begins with an approximate Hamiltonian $H_{0}$ composed of a complete orthogonal set of excitations, bosons or fermions, whose exact statistical mechanics is known. $H_{0}$ is complemented with an interaction Hamiltonian $H_{1}$, expressing the key effect missing from $H_{0}$, and often but not always treated as  a perturbation. The approach was developed to treat the wide variety of physical behaviors observed in condensed matter, and is well illustrated in the monographs of Pines~\cite{Pines_1997_MBP,Pines_1997_EES} and Kittel.~\cite{Kittel1963} The principles are evident in Boltzmann's theory for  a gas of  freely moving atoms
which interact via collisions,~\cite{Boltzmann} and in the theory of Born and coworkers for a crystal of harmonic phonons interacting via anharmonicity.~\cite{Born-vonKarman,Born-Huang}

In the absence of such a many body formulation, liquid dynamics theory has been advanced by a series of conceptual developments. An important early step was learning how to construct physically realistic interatomic potentials for nearly-free-electron metals from pseudopotential perturbation theory.~\cite{Harrison, Ashcroft, Heine-Weaire, Ashcroft-Stroud} It was shown that MD calculations using these potentials  give an excellent account of experimental data for elemental condensed systems, for example: for thermodynamic properties of crystals~\cite{DCW-PR1968,*DCW-PRB1970} and liquids,~\cite{B2} and for the liquid structure factor~\cite{B1,B3} and dynamic structure factor.~\cite{B3,C1,C2} 
\emph{Ab initio} MD was introduced,~\cite{D1,D2,D3} and has since become the method of choice for reliable calculations on many liquid types. 
The resulting physical picture of liquids as nuclei moving on the ground state adiabatic potential, and subject to conventional statistical mechanical averaging, is the basis of our theoretical work. Formal derivation of the corresponding condensed matter Hamiltonian is reviewed in Ch.~1 of Ref. \onlinecite{SPCL}.

The conceptual contribution of V-T theory is to classify the valleys that comprise the liquid's many-body potential surface as random or symmetric. The random valleys all have the same statistical mechanical averages, and together they dominate the potential surface in the thermodynamic limit.~\cite{DCW_PR1997_4179}  Originally a hypothesis, this classification has been numerically verified for various representative systems.~\cite{DCW_PR1997_4179,Wallace_Clements_1999,Clements_Wallace_1999,GDLV-DCW-PRE2007,Holmstrom2009,H}  The atomic motion is vibrations in one (any) random valley, interspersed by transits, which carry the system between random valleys. Accordingly, we define the extended random valley as the harmonic extension to infinity of any random valley, and take the Hamiltonian $H_{vib}$ for motion in one extended random valley as our zeroth order liquid Hamiltonian. To study any particular function, we first calculate the function assuming the Hamiltonian is $H_{vib},$ and then we examine the (often small) remainder to see how models of transit motion can account for it. (Studies of liquid motion in terms of hops between potential valleys date back at least to Stillinger and Weber~\cite{SW} and Zwanzig,~\cite{Zwanzig} but without the benefit of the V-T classification, which makes clear the choice of initial Hamiltonian.)
 
Transits have been observed in MD calculations for liquid Na and Ar, at very low temperatures where single transits are well resolved.~\cite{M} These transits occur in the highly correlated motion of a small local group of atoms. At and above melting, transits proceed at a high rate throughout the liquid.
In its initial formulation, V-T theory incorporated an empirical melting-entropy constant to represent the transit contribution to the entropy of elemental liquids.~\cite{DCW_PR1997_4179} This formulation requires extension in two ways: in order to treat \emph{all} thermodynamic properties, the transit entropy theory must be replaced by a free energy theory; and in order to make a purely \emph{liquid} theory, all sensitivity to the nature of the crystal or the melting process must be removed. These extensions were carried out through an analysis of the temperature dependence of experimental entropy for elemental liquids,~\cite{F} plus a statistical mechanics free energy model calibrated to the entropy results.~\cite{G} \emph{A priori} density functional theory (DFT) calculations then verified the theory to high accuracy for thermodynamic properties of liquid Na and Cu.~\cite{Bock6_PRB82_144101}

In nonequilibrium problems, V-T theory has achieved success in two applications to dynamic response. First, it was found that an \emph{a priori} calculation of the Brillouin peak dispersion curve, based on the vibrational motion alone, is in essentially perfect agreement with MD calculations and with experimental inelastic scattering data for liquid Na.~\cite{I} Second, in comparison with the benchmark theories of generalized hydrodynamics and mode coupling, a near-\emph {a priori} theory of self dynamic response was found to have significantly improved analytic properties and modestly improved accuracy.~\cite{us_Fqtself}
The purpose of the present work is to test the viability of V-T theory for the distinct part of the density autocorrelation function.

Hansen and McDonald~\cite{HMcD_3rded} define time correlation functions related to self and distinct contributions to dynamic response, and discuss the self dynamic structure factor, including its Gaussian approximation. These authors also point out it is possible in principle to measure separately the self and distinct parts. We have not found studies dedicated to the distinct part, presumably because the self part and the total function cover the entire theoretical problem. However, the distinct part must be studied separately, because the strong difference between self and distinct correlations implies a similar difference in their decorrelation processes.

Formulation of dynamic response theory is a quintessential quantum mechanics problem, analyzed for crystals by Maradudin and Fein,~\cite{Maradudin}  Ambegaokar, Conway and Baym,~\cite{Ambegaokar}  Cowley,~\cite{L1,L2} and Ashcroft and Mermin,~\cite{Ashcroft_Mermin} and for crystals and liquids by Lovesey~\cite{Bloch} and Glyde.~\cite{Glyde_book} However, our focus is the elemental liquids in general, so we omit the few quantum liquids and work in classical statistical mechanics. We study the density autocorrelation function $F(q,t)$, also called the intermediate scattering function,  for all wave vectors $q$ and time $t$. Expansions in powers of scattering events are inefficient, and we must work with the full theory, correct to all powers of $q$ (see Appendix N of Ref. \onlinecite{Ashcroft_Mermin}). We apply the primitive-lattice harmonic crystal analysis from the above references to a system moving in an extended random valley, then extract the classical limit to obtain the zeroth order liquid function $F_{vib}(q,t)$ (details may be found in Ref. \onlinecite{ET_Curr}). In the present paper we begin with the distinct part $F_{vib}^{d}(q,t)$, and introduce the decorrelating effects of transits to damp 
$F_{vib}^{d}(q,t)$ to the function representing V-T theory, $F_{VT}^{d}(q,t)$.

Sec.~II sets out the ``standard plan" for 
constructing the theoretical function $F_{VT}^{d}(q,t)$.
The perfect vibrational system provides 
equations for $F_{vib}^{d}(q,t)$ and
for the motional and structural correlations. Transits are responsible for all damping beyond the dephasing already contained in $F_{vib}^{d}(q,t)$. The transit contribution to $F_{VT}^{d}(q,0)$  is assigned to the structural correlation and is calibrated from MD. A crossover time $t_{c}(q)$ is defined, such that motional correlation damps to zero on $ 0 \leq t \leq t_{c}$, and structural correlation  damps to zero on $t \geq t_{c}$.

In Sec.~III, the massively averaged motion due specifically to transits on $ 0 \leq t \leq t_{c}$ is modeled as the transit drift, and a  theory is made for its contribution to motional decorrelation.  In Sec.~IV, the transit random walk theory from self dynamic response~\cite{us_Fqtself} is extended to the structural decorrelation. The agreement of theory with MD is at the remarkable level of $0.01|F_{MD}^{d}(q,0)|$, for all $t$, and all $q$ for which the standard plan applies. Sec.~V examines the validity of the standard decorrelation plan as function of $q$. 

In Sec.~VI, we review our main accomplishment: an expression $F_{VT}^d(q,t)$ that includes the physical ideas behind V-T theory. We also summarize the physical nature of the standard plan and the current status of transit modeling, and identify the path to making the present theory fully \emph {a priori}.

In this work, we frequently use MD results to validate or parametrize our theory. Our computational system represents liquid Na at 395 K, a bit above the melting temperature $T_{m}=371$~K. The system is a cube containing $N=500$ atoms, with periodic boundary conditions. The MD time step is 7~fs. The interatomic potential is based on pseudopotential theory,~\cite{Harrison} and has produced excellent agreement with a wide range of experimental data.~\cite{SPCL}
\section{Outline of the theory}
 
 We study the distinct  autocorrelation function,~\cite{fqtd} defined by
\begin{equation} \label{eq1}
F^{d}(q,t) = \frac{1}{N}  \sum_{K\neq L} \left< e^{-i {\bf q} \cdot \left( {\bf r}_{K}(t) - {\bf r}_{L}(0) \right) } \right>  ,
\end{equation}
for a system of atoms $K=1,...,N$, located at ${\bf{r}}_{K}(t)$ at time $t$. The average may be evaluated analytically for simple systems (like the vibrational system below) or numerically for a single MD system, on an equilibrium trajectory, and includes an average over the star of ${\bf q}$ when periodic boundary conditions are used (a star is the set of all wave vectors related by the cubic point group). The function measures static pair configurational correlations in terms of the structure factor $S^{d}(q) = F^{d}(q,0)$, and measures their statistical average time decay at $t>0$.

 \subsection{Perfect vibrational system}
 
To evaluate Eq.~(\ref{eq1}) for the vibrational system, we write 
\begin{equation}
{\bf r}_{K}(t) = {\bf R}_{K} + {\bf u}_{K}(t),
\label{eq2}
\end{equation}
where ${\bf{R}}_{K}$  is the random valley equilibrium position (structural site) and ${\bf{u}}_{K}(t)$ is the motional displacement. 
The vibrational average is then~\cite{ET_Curr}
\begin{equation} 
F_{vib}^{d} (q,t)  = \frac{1}{N} \sum_{K \neq L}  e^{-i {\bf q} \cdot {\bf R}_{KL}}~e^{-W_{K}({\bf q})} ~e^{-W_{L}({\bf q})}  
  ~e^{\left< {\bf q} \cdot {\bf u}_{K}(t)~ {\bf q} \cdot {\bf u}_{L}(0) \right >_{vib}},
\label{eq3}
\end{equation}
where ${\bf{R}}_{KL} = {\bf{R}}_{K}-{\bf{R}}_{L}$, $W_{K}({\bf q}) $ is the Debye-Waller factor,
\begin{equation}
W_{K}({\bf q}) = \frac{1}{2} \left< ({\bf q} \cdot {\bf u}_{K})^{2} \right >_{vib},
\label{eq4}
\end{equation}
and the motional  time-correlation functions are given by
\begin{equation} 
 \left< {\bf q} \cdot {\bf u}_{K}(t) {\bf q} \cdot {\bf u}_{L}(0) \right >_{vib} =   \frac {kT}{M}   
\sum_{\lambda} ({\bf q} \cdot {\bf w}_{K \lambda})~ ({\bf q} \cdot {\bf w}_{L \lambda}) \frac{\cos \omega_{\lambda} t}{\omega_{\lambda}^{2}}.
\label{eq5}
\end{equation}
The normal modes are labeled $\lambda=1,...,3N-3$, the three zero-frequency modes being omitted, mode $\lambda$ has frequency $\omega_{\lambda}$, and eigenvector $\lambda$ has Cartesian vector ${\bf{w}}_{K\lambda}$ at atom $K$. The right side of Eq.~(\ref{eq3})  is also averaged over each ${\bf{q}}$ star.

To calibrate the vibrational Hamiltonian for liquid Na, we quench the Na computational system to a random structure,~\cite{DCW_PR1997_4179,Wallace_Clements_1999,Clements_Wallace_1999,GDLV-DCW-PRE2007,Holmstrom2009,H} and there evaluate the equilibrium positions and the vibrational frequencies and eigenvectors.~\cite{Wallace_Clements_1999,Clements_Wallace_1999,GDLV-DCW-PRE2007}  These Na parameters are then used to evaluate the vibrational functions in this paper, and all other vibrational functions for liquid Na. Any one random structure is suitable for this purpose because of the statistical similarity of the random valleys in the thermodynamic limit.

Equation~(\ref{eq3})  for $F_{vib}^{d}(q,t)$ makes use of three data sets, the structural positions ${\bf{R}}_{K}$, the Debye-Waller factors in Eq.~(\ref{eq4}), and the motional time correlation functions in Eq.~(\ref{eq5}). These data sets are all strongly coupled inside the sums in Eq.~(\ref{eq3}), and that equation admits of no  acceptable decoupling approximation. The key to analysis is in the time dependence. 

$F_{vib}^{d}(q,t)$ is subject to natural vibrational decorrelation, or vibrational dephasing, of the $\cos \omega_{\lambda}t$ factors in Eq.~(\ref{eq5}), as $t$ increases from zero. This process is always present, and has the effect of reducing the time correlation functions to zero as $t \to \infty$. What remains of $F_{vib}^{d}(q,t)$ as $t \to \infty$ is therefore given by 
\begin{equation}
F_{vib}^{d} (q,\infty) = \frac{1}{N} \sum_{K \neq L}   e^{-i {\bf q} \cdot {\bf R}_{KL}}~e^{-W_{K}({\bf q})} ~e^{-W_{L}({\bf q})} .
\label{eq6}
\end{equation}
Because  the Debye-Waller factors are positive and increasing with $q$, Eq.~(\ref{eq4}), $F_{vib}^{d}(q,\infty) \to 0$ as $q$ increases.

$F_{vib}^{d}(q,t)$ accounts for all correlations in a perfect vibrational system. These correlations can be classified as motional and structural. We shall follow this intuitive notation, in a modification designed for V-T theory.
\begin{enumerate} [a)]
\item Motional correlation is that contained in the time correlation functions with $K \neq L$ in Eq.~(\ref{eq3}). Motional correlation is in the normal mode motion and resides in the eigenvectors, Eq.~(\ref{eq5}).
\item For vibrational motion in a single random valley, each atom $K$ remains within a small volume around its equilibrium position ${\bf R}_{K}$. This constraint describes the structural correlation, as the term is used here, and  it is contained in Eq.~(\ref{eq6}). 
\end{enumerate}

\subsection{Transit contribution to the structure factor}

\begin{figure} [h!]
\includegraphics [width=0.45\textwidth]{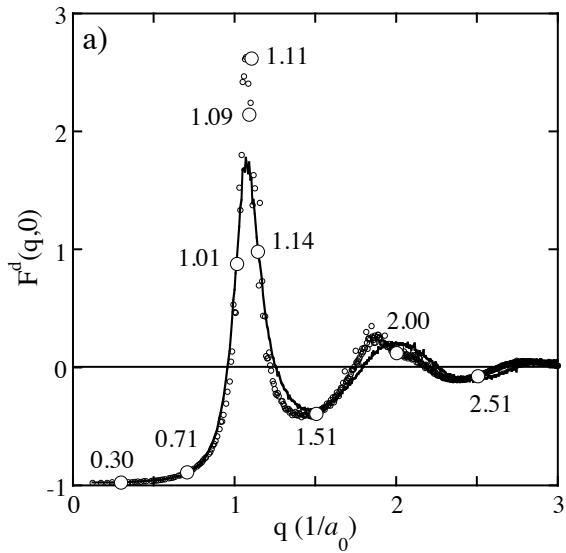}
\caption{Circles are $F_{vib}^{d}(q,0)$ at allowed $q$ (large circles identify the test $q$ from Table I) and the line is $F_{MD}^{d}(q,0)$.}
\label{fig2}
\end{figure}

Fig.~\ref{fig2} compares the distinct functions $F_{vib}^{d}(q,0)$  and $F_{MD}^{d}(q,0)$, which describe the initial correlations for all the allowed $q$ in our system.  Fluctuations are larger in the vibrational curve because it is calculated for a single random valley while the MD data averages over random valleys. The difference
\begin{equation}
A(q) = F_{MD}^{d}(q,0) - F_{vib}^{d}(q,0)
\label{eq7}
\end{equation}
is formally identified in V-T theory as the transit contribution.

As the figure shows, the vibrational contribution alone produces the correct peak structure, and it correctly locates the first peak. It also gets the width of the peak correct and provides about two thirds of its height. In fact, the transit contribution $A(q)$ is a small correction everywhere except the first peak tip, where it is large and negative. (We found a similar situation with the Brillouin peak dispersion curve~\cite{I}: the vibrational part alone gets the location of the peak in $S(q,\omega)$ but not its value.) While we do not yet have a microscopic theory of how transits affect initial correlations, the fact that $A(q)$ is large only at the nearest neighbor distance suggests that it is structural in the sense defined above and is thus contained in $F_{vib}^{d}(q,\infty)$. Accordingly we write
\begin{equation}
A(q) = C(q) F_{vib}^{d}(q,\infty),
\label{eq8}
\end{equation}
where $C(q)$ has no dependence on  $\left \{ {\bf R}_{K}\right \} $.

The formation of excess correlation by transits is one matter, but the transit-induced process that cause those correlations to decay is another. We do have a microscopic theory for those processes, and we begin to lay the groundwork for that theory in the next section.

\begin{center}
\begin{table*}[ht] 
\caption{\label{table1}MD and vibrational data for nine $q$ chosen as test cases in the present study. All functions are defined in text and equations.}
\begin{ruledtabular}
\begin{tabular}[c]{rrrrrlccc}
$q~(a_{0}^{-1})$ & $F_{MD}^{d}(q,0)$ & $ F_{vib}^{d}(q,0)$  & $A(q)$ & $ F_{vib}^{d}(q,\infty)$ & $t_{c}(q)$ (ps) & $s(q,t_{c}) (a_{0}$) & $B(q)$ & $q$-regime \\
\hline
0.29711 & -0.9730 & -0.9747   & 0.0017       & -0.9388      &  0.2937    &    0.761  &  1.286    & Brillouin peak \\ 
0.70726 & -0.8943 &  -0.8878  & -0.0065      & -0.7118      &  0.2595    &    0.774  &  1.265    & Brillouin peak \\ 
1.01482 & 0.9423  &  0.8787  & 0.0636        & 0.4921       &  0.3547    &    0.658  &   1.111   & first peak \\
1.09165 & 1.6937     &  2.1415   & -0.4478      & 1.6819        & 0.3336     &   0.613   &  1.405   &  first peak \\
1.10505 & 1.5504     & 2.6187    & -1.0683      & 2.1902        & 0.3222     &   0.528   &  1.436    &  first peak \\
1.14429 & 1.0201     & 0.9820    & 0.0381     & 0.6619        & 0.2787     &   0.844   &  1.564    &  first peak \\
1.50523 & -0.3540    & -0.3929   & 0.0389       &-0.1791        & 0.2790     &  0.583   &  1.134     & large $q$  \\ 
2.0041 & 0.2055        & 0.1227    & 0.0828       & -0.0021       & 0.198       &                &                 &  large $q$\\
2.5064 &  -0.1042      &  -0.0725 &  -0.0317     & -0.0134       &  0.153      &                &                 &  large $q$   \\
\end {tabular}
\end{ruledtabular}
\label{Table1}
\end{table*}
\end{center}

\subsection{Standard decorrelation plan}

We chose a representative set of $q$ values to use in developing  the present theory.  These $q$ are listed in Table~\ref{Table1}, along with the theoretically important functions for each $q$. The nine $q$ in Table~\ref{Table1}  are among the $17~q$ for which self-decorrelation calculations were done.~\cite{us_Fqtself}

The vibrational density autocorrelation function, normalized to MD data at $t=0$, is $F_{vib}^{d}(q,t) + A(q)$. Our object is to construct a theory for the transit-induced decorrelation of this function.

\begin{figure} [h!]
\includegraphics [width=0.35\textwidth]{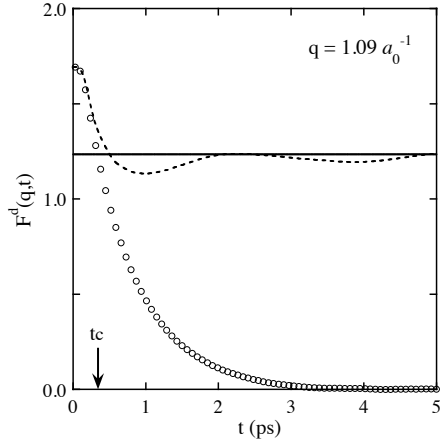}

\caption{Circles are $F_{MD}^{d}(q,t)$,  dashed line is $F_{vib}^{d}(q,t) + A(q)$, and solid line is   $F_{vib}^{d}(q,\infty) + A(q)$. $t_{c}$ is the time when MD crosses the horizontal solid line. The vibrational excess is $F_{vib}^{d}(q,t) - F_{vib}^{d}(q,\infty)$ at $t \geq t_{c}$. }
\label{fig3}
\end{figure}

Fig.~\ref{fig3} shows curves of $F_{MD}^{d}(q,t)$ and $F_{vib}^{d}(q,t) + A(q)$ for $q=1.09$, at the tip of the first peak. The curves agree at $t=0$, but the MD curve damps faster, and falls increasingly below the vibrational curve as $t$ increases. Recall the vibrational curve contains natural vibrational decorrelation, which makes the curve converge eventually to the constant $F_{vib}^{d}(q,\infty)+A(q)$. Through the convergence process, from around $0.4$~ps to around $5$~ps, the vibrational curve displays the vibrational excess, an oscillatory variation about its $t \to \infty$ limit. The feature is due to very-slowly-damped lowest frequency normal modes. The same vibrational excess is present in all the $F_{vib}^{d}(q,t)$ curves, and in the curves of  the self autocorrelation function $F_{vib}^{s}(q,t)$ as well. Its appearance in  $F_{vib}^{s}(q,t)$ was noted previously.~\cite{us_Fqtself} Exceptionally, we note that the vibrational excess does not appear at sufficiently large $q$, in $F_{vib}^{d}(q,t)$ nor in $F_{vib}^{s}(q,t)$, because the functions converge to zero before the excess develops.

\begin{figure} [h!]
\includegraphics [width=0.40\textwidth]{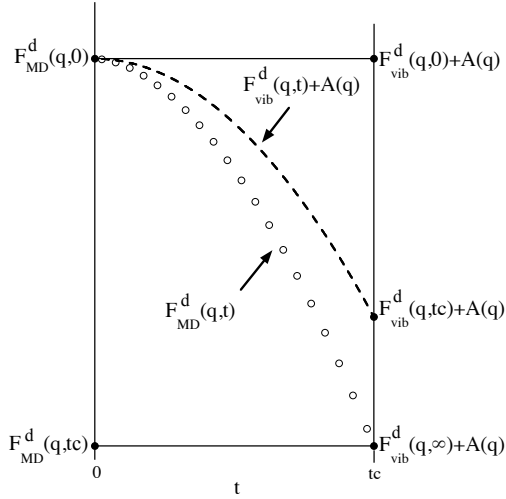}
\caption{
The same curves that are plotted in Fig.~\ref{fig3}  during the motional decorrelation period ($t \leq t_{c}$). The decrease in the vibrational curve is due to vibrational dephasing. Transit-induced decorrelation is supposed to damp the vibrational curve to the MD curve.} 
\label{fig4}
\end{figure}

In fact, the vibrational excess cannot be present in the liquid state, because the long-time normal-mode correlation cannot survive in the presence of transits. The vibrational excess must therefore be damped out of $F_{vib}^{d}(q,t)$ by transit-induced motional decorrelation. The situation is shown graphically in Fig.~\ref{fig4}. The constant $F_{vib}^{d}(q,\infty) +A(q)$  is the value of the normalized vibrational autocorrelation function with all motional correlation damped and all structural correlation present. The figure suggests, and we shall adopt, a simplifying approximation for the complete decorrelation process: make the transit-induced motional and structural decorrelations sequential, the first ending and the second beginning at the same (crossover) point.   Then with  the crossover time denoted $t_{c}(q)$, the theory must satisfy
\begin{equation}
F_{VT}^{d}(q,t_{c}) = F_{vib}^{d}(q,\infty) + A(q).
 \label{eq9}
 \end{equation}
 As for $t_{c}(q)$ itself, we shall calibrate it with the help of MD data by
\begin{equation}
 F_{MD}^{d}(q,t_{c}) =  F_{vib}^{d}(q,\infty) + A(q),
 \label{eq10}
 \end{equation}
 as shown in Fig.~\ref{fig4}.  The standard decorrelation plan is then described as follows.

 \begin{itemize}
 \item  Introduce transit-induced motional decorrelation to damp the vibrational curve toward the MD curve on $0 \leq t \leq t_{c}(q)$.
 \item  Introduce transit-induced structural decorrelation to damp the line $(F^{d}_{vib}(q,\infty) +A(q))$ toward the MD curve on $t \geq t_{c}(q)$.
\end{itemize}

In the process of this study we shall find those $q$ for which the standard plan applies, and shall summarize the results in Sec. V. While such a qualitative conclusion can be drawn with some reliability, quantitative behavior differs with $q$. In this work, one should be mindful that every $q$ measures a different correlation.

\section{Theory for transit-induced motional decorrelation}

\subsection{Transit drift}

For the present construction, we start with a perfect  vibrational system, for which the equilibrium trajectory has perfect vibrational configurations. Vibrational dephasing operates in 
$F_{vib}^{d}(q,t)$, and transits will contribute additional damping of the motional correlations. The decorrelation must begin from zero at any time chosen for the start of the calculation. This condition is satisfied by the vibrational dephasing, but it must be made an initial condition for the transit-induced decorrelation.

\begin{figure} [h!]
\includegraphics [width=0.35\textwidth]{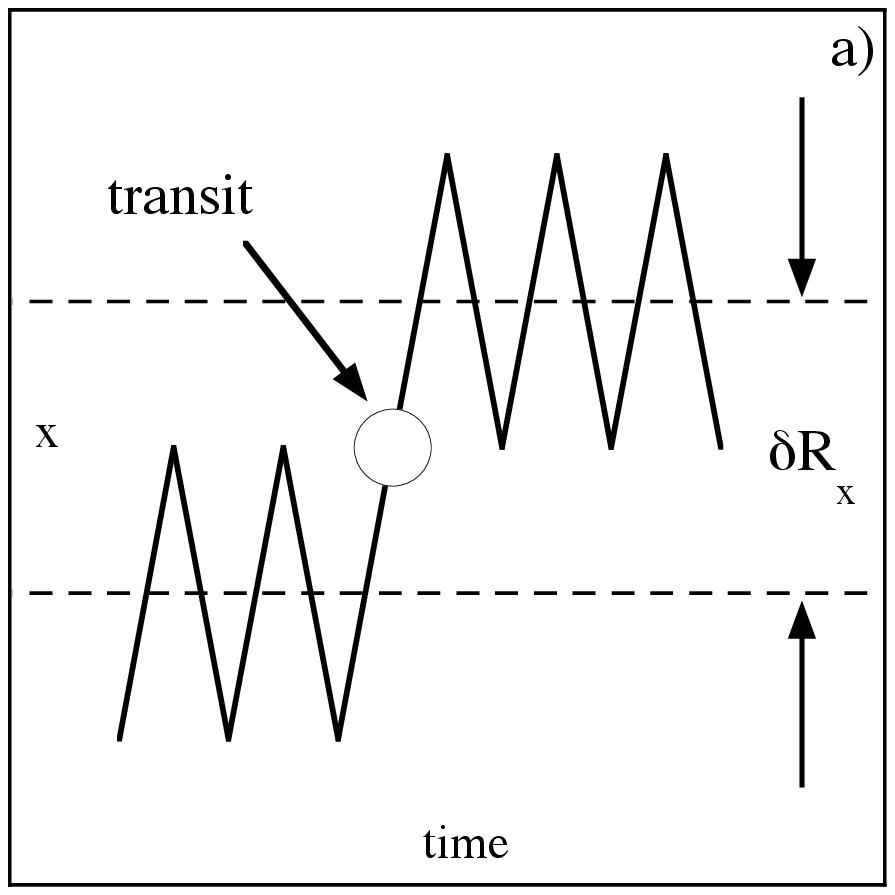}
\includegraphics [width=0.35\textwidth]{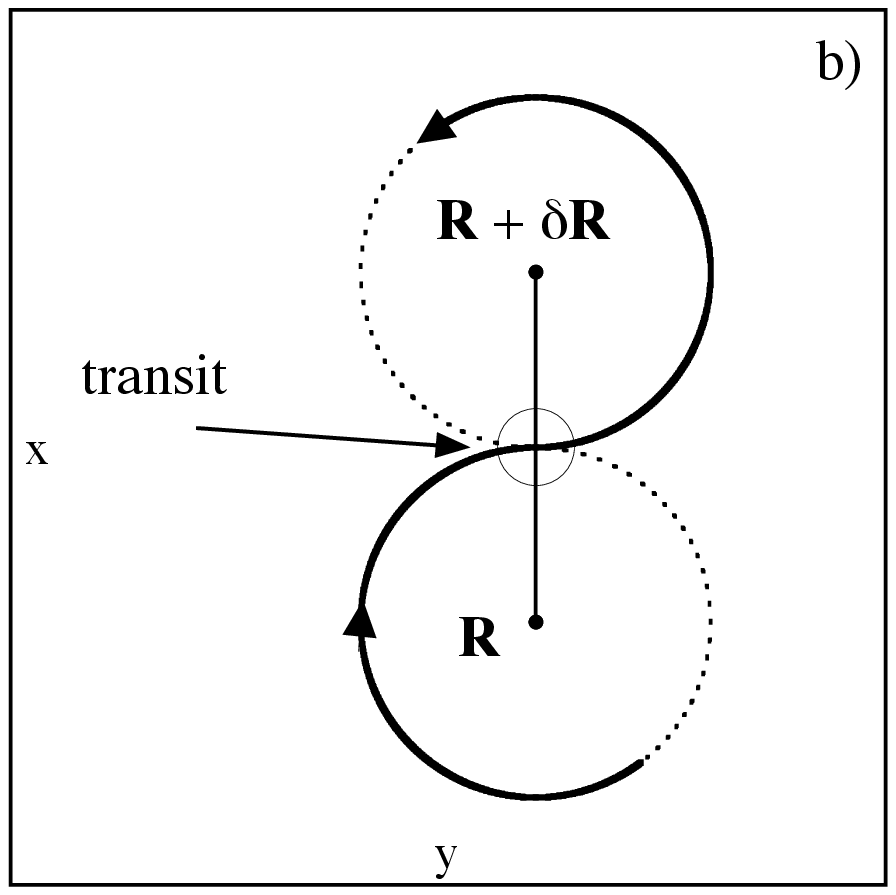}
\caption{Representation of Cartesian coordinates for a transiting atom. a) shows $x~vs~t$ for one (any) of a local group of atoms that move together in a single transit (from Ref.~\onlinecite{M}). b) is a planar model of the motion in a).}
\label{fig8}
\end{figure}

While the transits observed in MD calculations occur in correlated groups,~\cite{M} the present statistical mechanics theory, Eq.~(\ref{eq1}), requires only the separate motion of one atom at a time. In the MD transits,
 the single atom motion shows a simple uniform behavior in graphs of the atomic Cartesian coordinates as functions of time. A schematic representation of one such graph is shown in Fig.~\ref{fig8}a. A model representation of the same motion in the $x-y$ plane is shown in Fig.~\ref{fig8}b. The motion in 3-d is minimally described as follows.

\begin{enumerate}[a)]
\item  Before transit, the atom is in motion on the vibrational surface, approximately a sphere, about equilibrium position ${\bf R}$.
\item After transit, the atom is in motion on the vibrational surface, approximately the same sphere, about equilibrium position ${\bf R} + \delta{\bf R}$.
\item The transit itself is merely the crossing of the boundary between two potential energy valleys, and is essentially instantaneous.
\end{enumerate}
Because the transit is instantaneous, the complete motion as described is consistent with every equilibrium configuration being a perfect vibrational configuration.

Two transit parameters, previously calibrated for liquid Na at $395$ K, are needed. $\delta R$ is the mean change in the equilibrium position of one atom in one transit, and is evaluated from the transits observed in MD.~\cite{M} $\nu$ is the mean transit rate for one atom, and is evaluated by fitting a transit random walk to MD data for the self diffusion coefficient.~\cite{us_Fqtself} The results, in the form needed here, are 
\begin{eqnarray} \nonumber
 \frac{1}{2} \delta R&  = & 0.88~a_{0},\\
 \nu^{-1} & = & 0.26 ~\text{ps},
 \label{eq14}
 \end{eqnarray}
 where $\nu^{-1}$ is the mean period between successive transits for one atom. For comparison, the nearest-neighbor distance is $7.0~a_{0}$,~\cite{Clements_Wallace_1999} and the mean vibrational period is $0.40$~ps.~\cite{R2}
 
 Consider a single atom trajectory containing one transit similar to that depicted in Fig.~\ref{fig8}, in three dimensions. We ask for the motion contribution resulting specifically from the transit. The transit location is the midpoint of $\delta{\bf R}$, and for all transits with a given $\delta{\bf R}$ the collective motion has cylindrical symmetry about $\delta{\bf R}$. Hence the mean transit-induced motion, ${\bf s}(t)$, is along $\delta{\bf R}$, from the transit location, to the ultimate mean position $\frac{1}{2} \delta{\bf R}$ away.
 
 Consider now the entire system. The motions ${\bf s}(t)$ are proceeding throughout the system at a constant rate.  For each spatial direction, the ${\bf s}(t)$ collectively produce a steady state motion  characterized in the mean per atom by a constant velocity. At any time, these motions are uniformly distributed over angles. The total motion, measured per atom, is referred to as the transit drift.
 
 The last theoretical issue is the state of the transit drift at the endpoint $t_{c}$ of the motional decorrelation period. We expect a single transit per atom, operating along with the natural dephasing, to damp the motional correlation to zero. This takes a time $\nu^{-1}$, hence provides a theoretical prediction for $t_{c}(q)$. The drift  magnitude $s(t_{c})$ achieved in time $t_{c}$ is close to $\frac{1}{2}\delta R$. The theory therefore makes three qualitative predictions about the microscopic process:
\begin{eqnarray} \label{pinco} \nonumber
 t_{c}(q)&  \approx & \nu^{-1},\\ 
 s(t_{c})& \approx &\frac{1}{2} \delta R,\\  \nonumber
 s(t) &\varpropto & t.
 \end{eqnarray}
 
 For the  theory of transit-induced motional decorrelation, we write
  \begin{equation}
 F_{VT}^{d}(q,t) = F_{vtr}^{d}(q,t) + A(q),~~~ 0 \leq t \leq t_{c}(q),
 \label{eq16}
 \end{equation}
 where $F_{vtr}^{d}(q,t)$ is $F_{vib}^{d}(q,t)$, Eq.~(\ref{eq3}), modified to include the transit-induced motion:
\begin{equation} 
F_{vtr}^{d} (q,t)  =  \frac{1}{N} \sum_{K \neq L}  e^{-i {\bf q} \cdot {\bf R}_{KL}}    \left < \left < e^{-i {\bf q} \cdot ({\bf u}_{K}(t)+{\bf s} _{K}(t)- {\bf u}_{L}(0)) } \right >_{tr} \right >_{vib},
\label{eq17}
\end{equation}
 where subscript \emph{tr}  denotes a transit property. Here the displacements ${\bf u}_{K}(t)$ and ${\bf u}_{L}(0)$ express normal-mode motion, while  ${\bf s}_{K}(t)$ represents the additional motion due to transits. We now decouple the   ${\bf s}_{K}(t)$, by evaluating them as the steady-state  average ${\bf s}(t)$, with the initial condition ${\bf s}(0)=0$. Then the factor $e^{-i {\bf q} \cdot {\bf s}(t)}$  can be separately averaged over angles and removed from the sums. The remaining factor is just $F_{vib}^{d}(q,t)$, so that 
  \begin{equation}
 F_{vtr}^{d}(q,t) =F_{vib}^{d}(q,t) \chi_{tr}^{d}(q,t),
 \label{eq18}
 \end{equation}
where the decorrelation factor (damping factor) is 
 \begin{equation}
 \chi_{tr}^{d}(q,t) =\frac {\sin {q s(t)}}{q s(t)}, ~~~ 0 \leq t \leq t_{c}(q).
 \label{eq19}
 \end{equation}
From Eqs.~(\ref{eq18}) and (\ref{eq19}), the two motional decorrelation processes, natural and transit-induced, operate independently, and they satisfy the condition of zero decorrelation at $t=0$. 

From Eqs.~(\ref{eq16}) and (\ref{eq18}), the  theory with motional decorrelation is
  \begin{equation}
 F_{VT}^{d}(q,t) = F_{vib}^{d}(q,t) \chi_{tr}^{d}(q,t)+ A(q), \quad \quad 0 \leq t \leq t_{c}(q).
 \label{eq20}
 \end{equation}
 We have $\chi_{tr}^{d}(q,0) =1$, and the endpoint magnitude is calibrated by comparing Eq.~(\ref{eq20}) with Eq.~(\ref{eq9}), to find
 \begin{equation}
 \chi_{tr}^{d}(q,t_{c}) =\frac {F_{vib}^{d}(q,\infty)}{F_{vib}^{d}(q,t_{c})}.
 \label{eq21}
 \end{equation}
 Then with Eq.~(\ref{eq19}),  $\chi_{tr}^{d}(q,t_{c})$ can be solved for $s(t_{c})$, and the third line in Eq.~(\ref{pinco}) prescribes
 \begin{equation}
 s(t)=\frac{t}{t_{c}} s(t_{c}).
 \label{eq22}
 \end{equation}
 
\subsection{Comparison of theory with MD}

\begin{figure} [h!]
\includegraphics [width=0.35\textwidth]{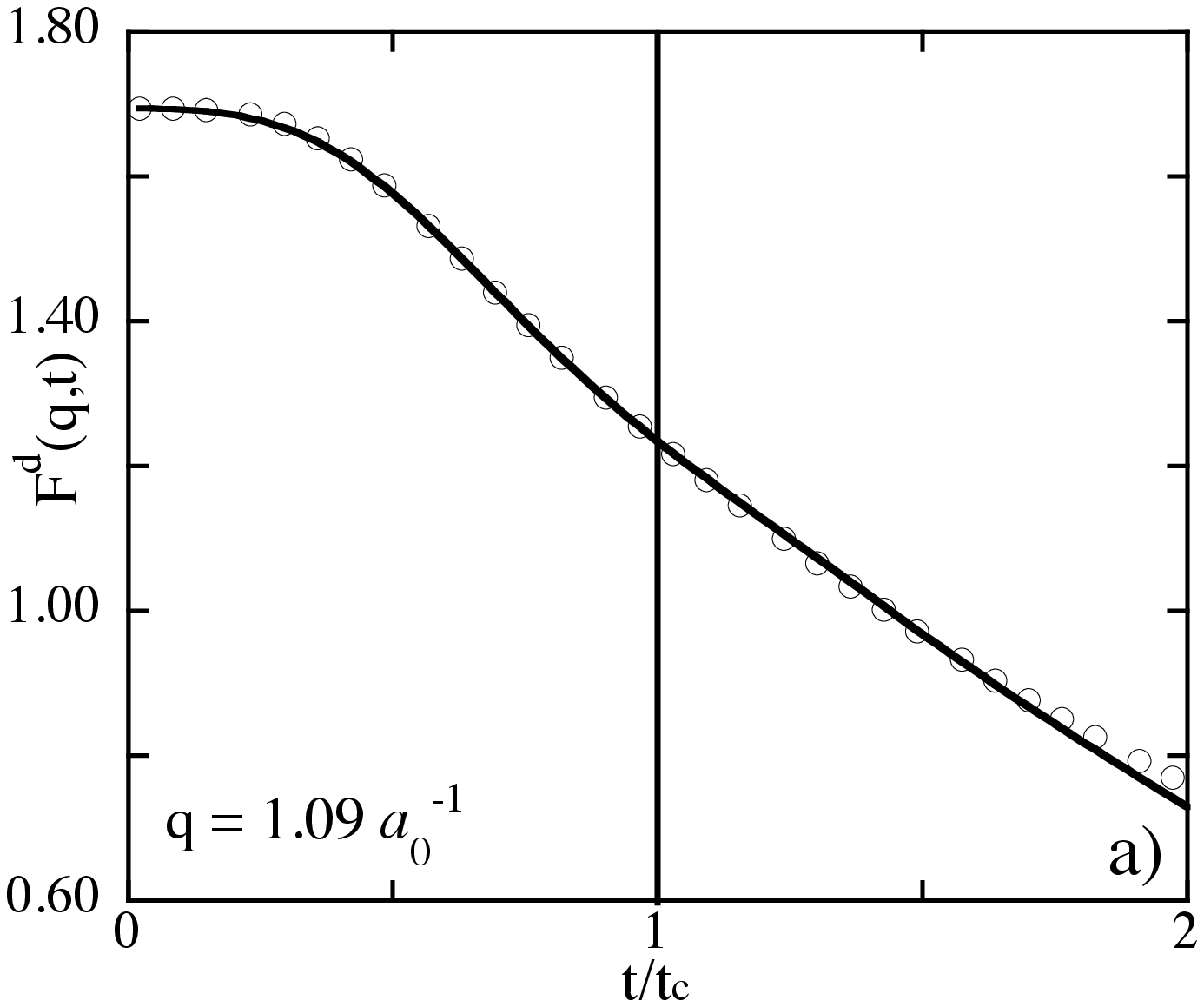}
\includegraphics [width=0.35\textwidth]{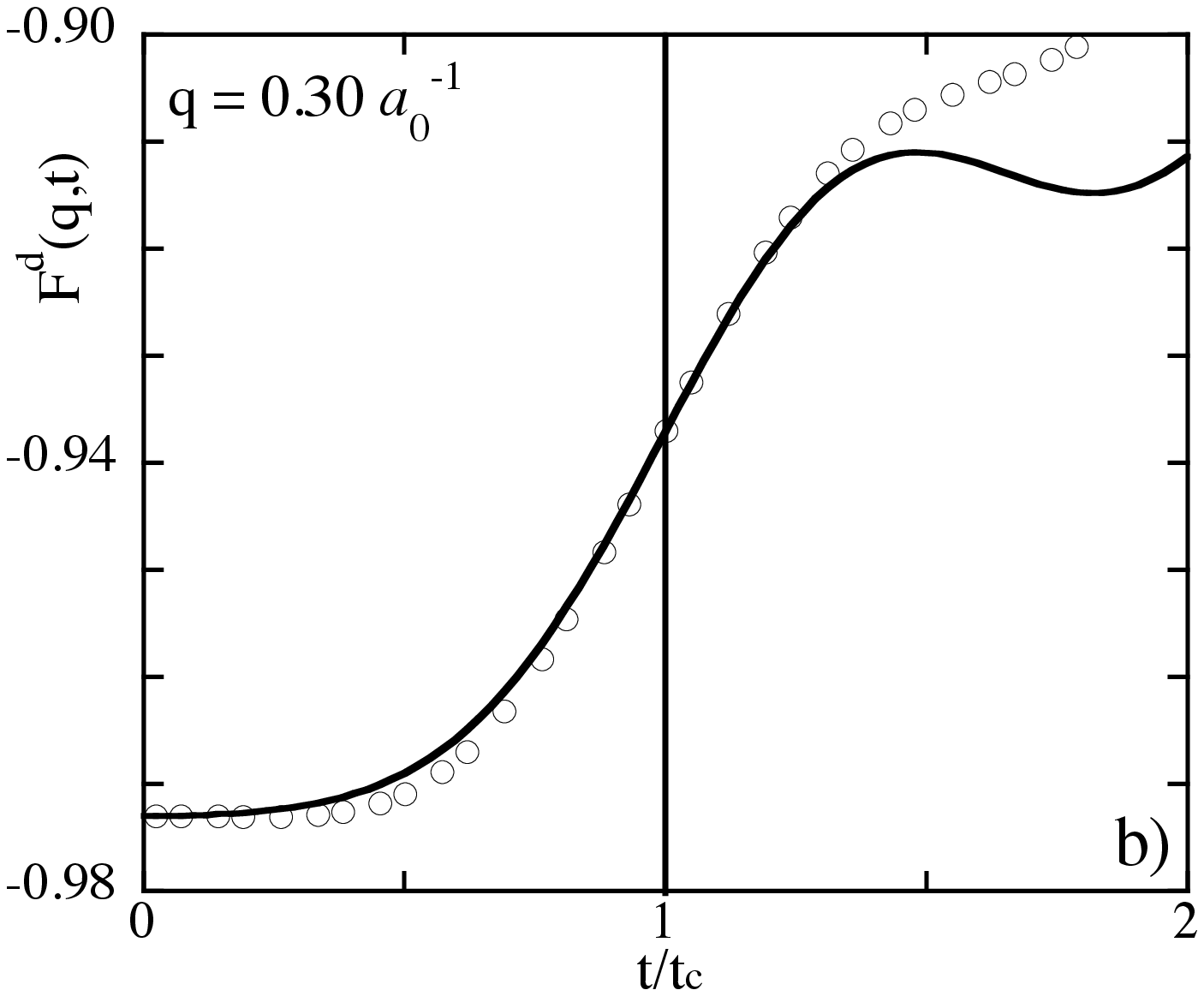}
\includegraphics [width=0.35\textwidth]{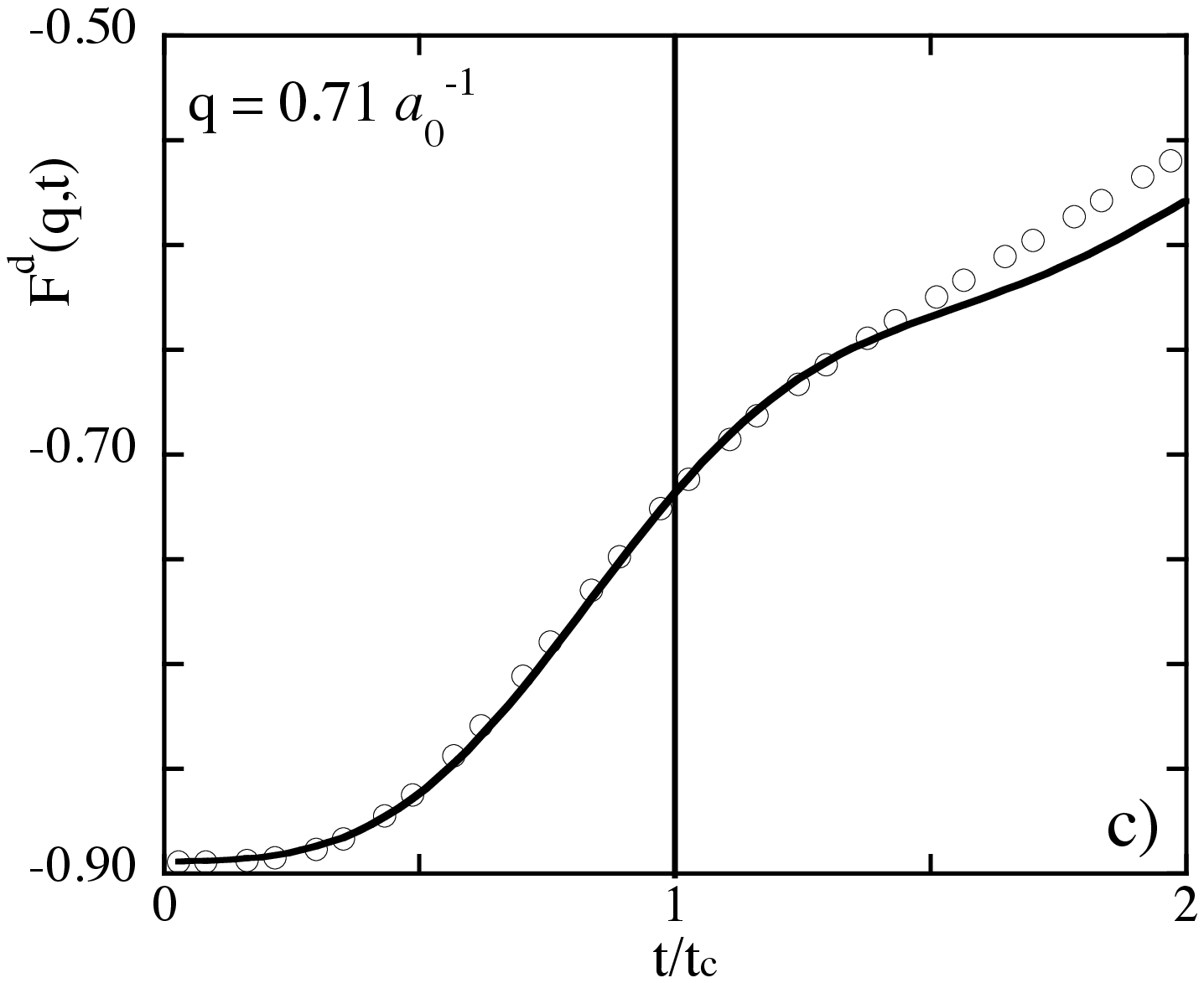}

\caption{For $F^{d}(q,t)$: Comparison of V-T theory (solid line) with MD (circles) on  $ 0 \leq t \leq 2t_{c}$. Theory is intended to apply on $ 0 \leq t \leq t_{c}$. a) is representative of all $q$ in the first peak regime; b) and c) are characteristic of the Brillouin peak regime.}
\label{fig567}
\end{figure}

Comparison of $F_{VT}^{d}(q,t)$ with $F_{MD}^{d}(q,t)$ is shown in Fig.~\ref{fig567}a for $q=1.09$ and $0\leq t \leq 2t_{c}$. On the intended theoretical range,  $0\leq t \leq t_{c}$, the agreement of V-T with MD is excellent. The time extension shows the theory is accurate to well beyond $t_{c}$, revealing the possibility of an improved theory in which the crossover is dispersed over a time interval. The same graphs for $q=1.01, 1.11 $ and $1.14$, all on the first peak, are similar to Fig.~\ref{fig567}a, with maximum V-T error up to $0.012$ on  $0\leq t \leq t_{c}$.

Comparison of $F_{VT}^{d}(q,t)$ with $F_{MD}^{d}(q,t)$ is shown for $q=0.30$ and  $0.71$ in  Figs.~\ref{fig567}b and \ref{fig567}c, respectively. For $q=0.30$ on  $0\leq t \leq t_{c}$, the maximum V-T error is $0.002$. For $q=0.71$ on $0\leq t \leq t_{c}$, the maximum V-T error is $0.006$. At $t > t_{c}$, for $q=0.30$ and $0.71$, the  theory provides some transit-induced damping of the Brillouin peak oscillation. However, as the figures show, the  damping is still insufficient at $t > t_{c}$.

\section{Theory for transit-induced structural decorrelation}

\subsection{Transit random walk}

In the period $0\leq t \leq t_{c}$, each atom undergoes one instantaneous transit, the ``first transit", and the subsequent motion specifically due to the transit is averaged to the steady-state transit drift. The drift has time and distance scales matching the normal mode motion, hence is efficient in decorrelating this motion, and does so until the decorrelation is complete at $t_{c}$. At the same time, the drift has no effect on structural correlation, because it does not noticeably alter the structure.

 At $t_{c} \leq t \leq 2t_{c}$, each atom undergoes its second transit after $t=0$. Here and at all later times, the mean transit effect is still the drift, which persists as the same steady flow, while the motional correlation remains zero. But the net drift of each atom accumulates, surpassing the time and distance scales of the normal mode motion, at which point the drift constitutes motion of the atomic equilibrium positions.  \emph{This} motion will damp the system structural correlation.
 
 The analysis now is specifically for $t\geq t_{c}$. When structural decorrelation begins, at $t_{c}$, Eqs.~(\ref{eq8}) and (\ref{eq9}) combine to yield
 \begin{equation}
 F_{VT}^{d}(q,t_{c}) = [1+C(q)] F_{vib}^{d}(q,\infty).
 \label{eq23}
 \end{equation}
 The structural information is explicit in the set $\{{\bf R}_{K}\}$ contained in  $F_{vib}^{d}(q,\infty)$. When the equilibrium positions begin to move, ${\bf R}_{K}$ becomes ${\bf R}_{K}(t)$, with ${\bf R}_{K}(t_{c})={\bf R}_{K}(0)$. The theoretical time dependence is then
 
 \begin{eqnarray} \nonumber
 F_{VT}^{d}(q,t) &= &\left [1+C(q)\right ] \frac{1}{N} \sum_{K \neq L} \left <   e^{   -i{\bf q} \cdot ({\bf R}_{K} (t) - {\bf R}_{L}(0))}  \right >_{tr} \\
 && \times~~e^{-W_{K}(q)} e^{-W_{L}(q)}.
 \label{eq24}
 \end{eqnarray}
 Denote the average $\langle \cdots \rangle_{tr}$ by $D_{KL}(t)$, suppressing the $q$ dependence. Consider a time interval $\delta t$ so small that atom $K$ is very unlikely to transit more than once in $\delta t$. Then in $\delta t$, $D_{KL}(t)$ changes by
  \begin{equation}
 \delta D_{KL}(t) = \left < \left [e^{-i {\bf q} \cdot {\bf R}_{K}(t+\delta t) } - e^{-i {\bf q} \cdot {\bf R}_{K}(t) } \right  ]  e^{i{\bf q} \cdot  {\bf R}_{L}(0)} \right >_{tr}.
 \label{eq25}
 \end{equation}
In $\delta t$, atom $K$ does not transit, or atom $K$ transits once with probability $\nu\delta t$. If atom $K$ does not transit, ${\bf R}_{K}(t+\delta t) = {\bf R}_{K}(t)$.  If atom $K$ does  transit, ${\bf R}_{K}(t+\delta t) = {\bf R}_{K}(t) + \delta {\bf R}_{K}$,
where $\delta {\bf R}_{K}$ has no time dependence. Then Eq.~(\ref{eq25}) becomes
  \begin{equation} 
 \delta D_{KL}(t)  =   \left < [e^{-i  {\bf q} \cdot \delta {\bf R}_{K}} - 1 ]   e^{   -i{\bf q} \cdot ({\bf R}_{K} (t) - {\bf R}_{L}(0))}   \right >_{tr} \nu \delta t.
 \label{eq26}
 \end{equation}

In a first approximation we can assume $|\delta {\bf R}_{K}| = \delta R$, the same for all transits, while the direction of $\delta {\bf R}_{K}$ is uniformly distributed and uncorrelated with ${\bf R}_{K}(t) - {\bf R}_{L}(0)$ in Eq.~(\ref{eq26}). In that case, the square bracket in Eq.~(\ref{eq26}) can be separately averaged over angles of $\delta {\bf R}$ to give 
  \begin{equation}
 \frac{\delta D_{KL}(t)}{\delta t} \approx -\gamma(q) D_{KL}(t),
 \label{eq27}
 \end{equation}
 where
  \begin{equation}
 \gamma(q) = \nu \left [1-\frac{\sin{q\delta R}}{q\delta R} \right ].
 \label{eq28}
 \end{equation}

 The above derivation was presented in our self decorrelation study,~\cite{us_Fqtself} where $\gamma(q)$  is found to provide accurate structural damping. Here, however, Eq.~(\ref{eq27}) contains an error due to the neglect of directional correlation in Eq.~(\ref{eq26}) between the vectors   $\delta {\bf R}_{K}$ and ${\bf R}_{K}(t)-{\bf R}_{L}(0)$. This  correlation is especially strong for the first transit and for atoms $K,L$ being nearest neighbors at $t=0$. The result is an anisotropic contribution to the random walk. The corresponding damping coefficient is $G(q)$, which we write in the form
  \begin{equation}
G(q) = \gamma(q) + \delta G(q),
 \label{eq29}
 \end{equation}
 where $\delta G(q)$ represents the anisotropic contribution to the random walk decorrelation. Then Eq.~(\ref{eq27}) becomes
   \begin{equation}
 \frac{\delta D_{KL}(t)}{\delta t} = -G(q) D_{KL}(t).
 \label{eq30}
 \end{equation}
 The equation integrates to
    \begin{equation}
D_{KL}(t) = {e^{-G(q)(t-t_{c})}} D_{KL}(t_{c}).
 \label{eq31}
 \end{equation}
 Putting this into Eq.~(\ref{eq24}) gives the structural decorrelation theory,
 \begin{equation}
 F_{VT}^{d}(q,t) = \left[ 1+C(q) \right ] F_{vib}^{d}(q,\infty) ~e^{-G(q)(t-t_{c}(q))}, \quad \quad t \geq t_{c}(q).
 \label{eq32}
 \end{equation}
Equation~(\ref{eq20}) for $t \leq t_c$ and Eq.~(\ref{eq32}) for $t \geq t_c$ form our complete theory for the time evolution of $F^d(q, t)$. 

\subsection{Comparison of theory with MD}

\begin{figure} [h!]
\includegraphics [width=0.35\textwidth]{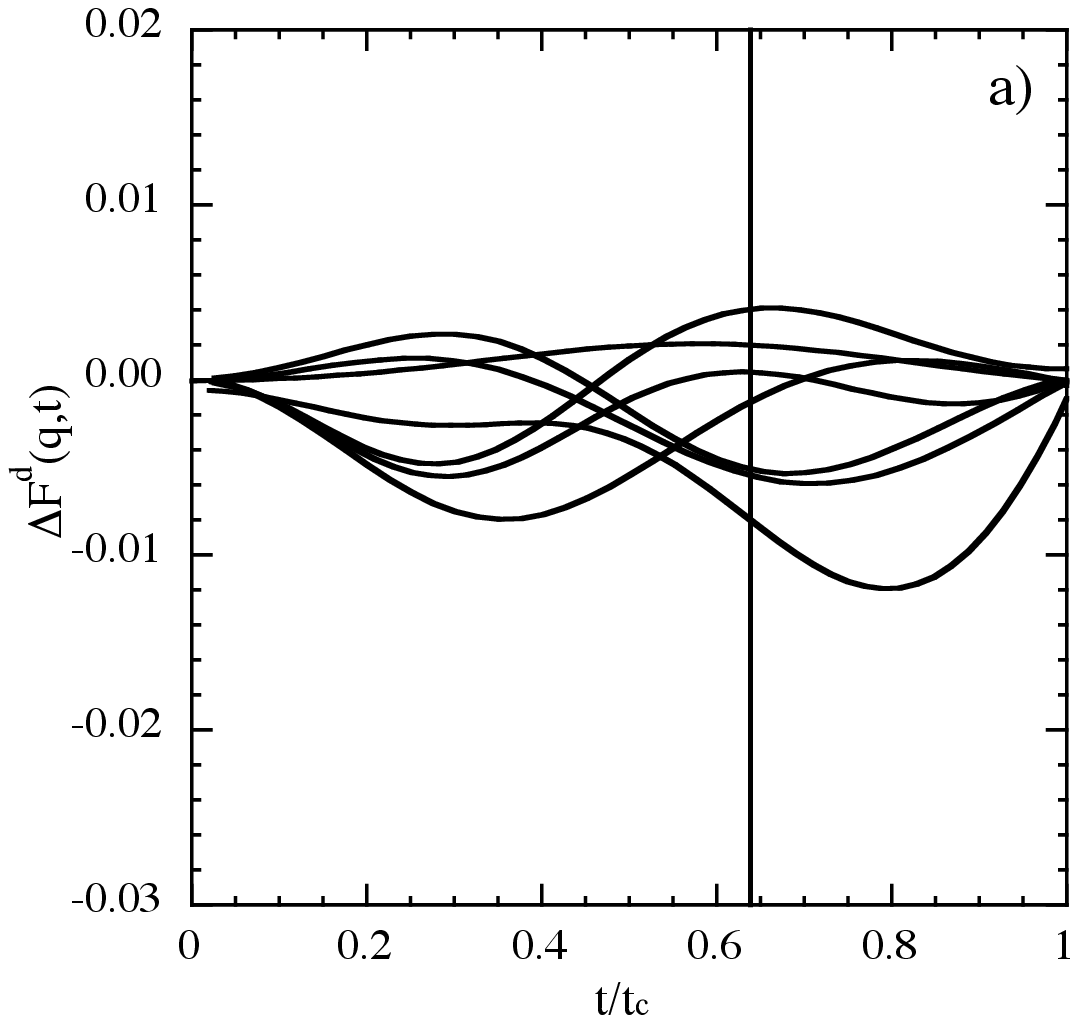}
\includegraphics [width=0.35\textwidth]{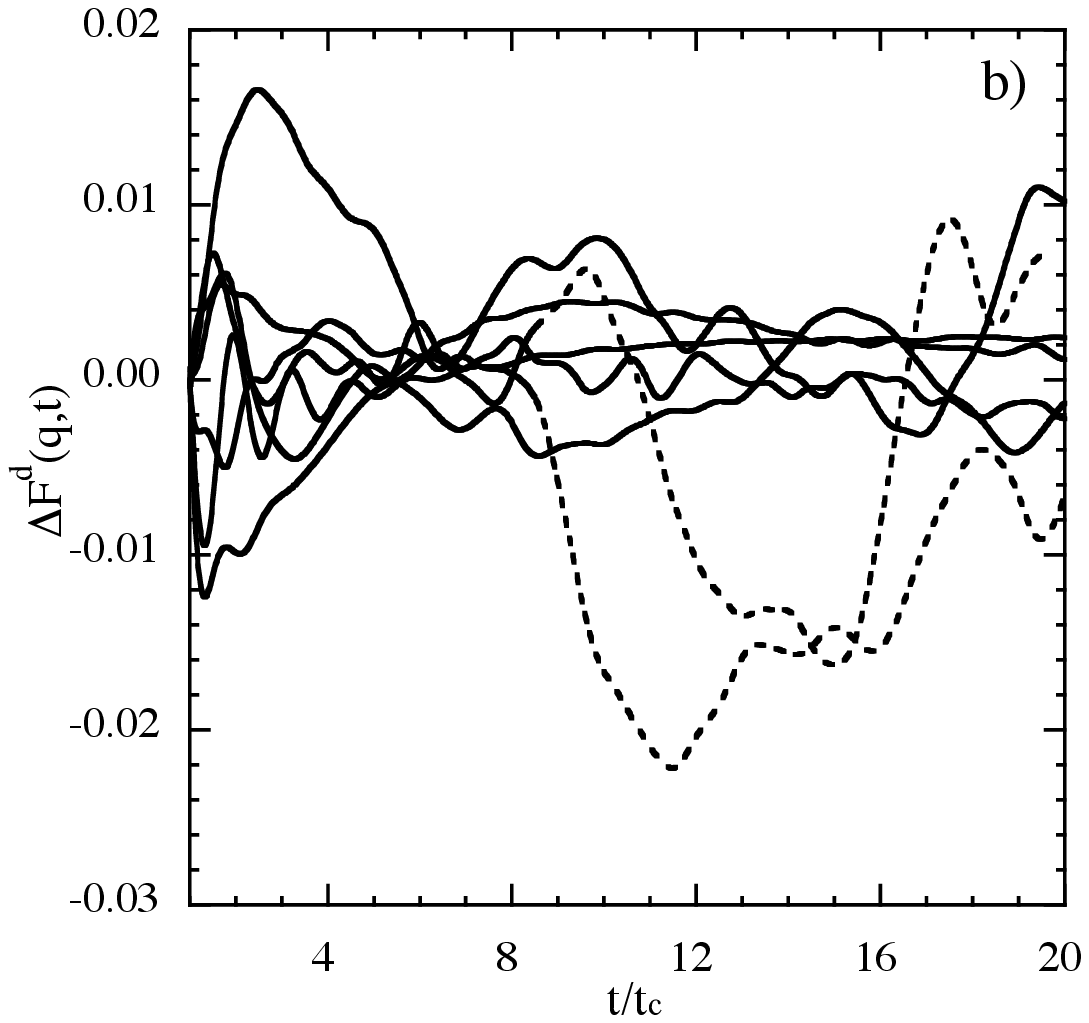}
\caption{(Color online) Deviation of  V-T theory from MD, Eq.~(\ref{eq33}), $vs$  $t/t_{c}$, for the seven $q$ where the standard plan is valid, see Table~II. a) motional decorrelation, 
 b) structural decorrelation, and the dotted portions show the MD long time tail (see Fig.~\ref{fig13}).
}
\label{fig9}
\end{figure}

Here we shall compare the full theory with MD, for $t\geq 0$. The functions needed to calibrate $F_{VT}^{d}(q,t)$ are listed in Tables~I and II. $G(q)$ is determined by fitting a straight line to $\log F_{MD}^{d}(q,t)$ vs $t$ on $t>t_{c}(q)$, where Eq.~(\ref{eq32}) is supposed to be valid. $\gamma(q)$ is from Table~I of Ref. \onlinecite{us_Fqtself}. For all test $q$ for which the standard plan applies, the deviations $\Delta F^{d}(q,t)$ are graphed in Fig.~\ref{fig9}, where
  \begin{equation}
 \Delta F^{d}(q,t) = F_{VT}^{d}(q,t) - F_{MD}^{d}(q,t).
 \label{eq33}
 \end{equation}

\begin{figure} [h!]
\includegraphics [width=0.35\textwidth]{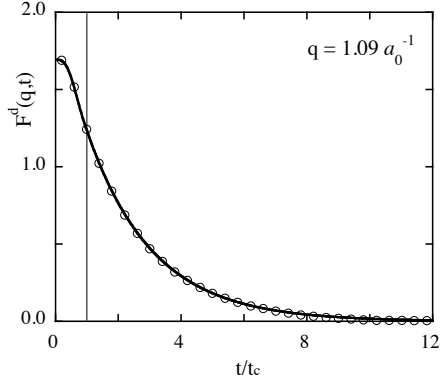}
\caption{For $F^{d}(q,t)$: Comparison of the complete  V-T theory (line) with MD (circles), for a $q$ representative of the first peak regime. Vertical line is at $t_c$.}
\label{fig10}
\end{figure}

Consider the first peak regime. The challenge here is the function $A(q)$, which is very large (and negative) at the tip of the first peak, $q=1.09$ and $1.11$, while it has the character of a small systematic correction at all other $q$, including the two additional $q$ in the first peak regime (see Table~I, also Fig.~\ref{fig2}). The significant result is that the theory accounts uniformly well for the entire decorrelation process for all four test $q$ in the first peak regime. The example of $q=1.09$ is shown in Fig.~\ref{fig10}. For the other three $q$, the overall shape is similar to Fig.~\ref{fig10}, and the comparison of theory with MD is qualitatively the same (Fig.~\ref{fig9}). 
Two details are specific to the first peak regime. First, $\delta G / \gamma$ is small and negative at the tip of the first peak ($q=1.09$ and $1.11$), $\delta G / \gamma$ is small and positive on the leading edge ($q=1.01$), and $\delta G / \gamma$ is negligible on the trailing edge ($q=1.14$). The physical implication will be discussed in Sec.~VI. Second, while the slope of theory is not controlled at $t_{c}$, so that a slope discontinuity is to be expected, there is  no significant discontinuity in the first peak regime (Fig.~\ref{fig10}). Note the time scale $t/t_{c}$ is an approximate count of the number of transits per atom from $t=0$.

\begin{table}
\caption{\label{table1}Data required for the structural decorrelation, at $t\geq t_{c}(q)$. $\gamma(q)$ is from Table I of Ref.~\onlinecite{us_Fqtself}.}
\begin{ruledtabular}
\begin{tabular}[c]{cccl}
$q~(a_{0}^{-1})$ & $1+C(q)$ & $ \gamma(q) $(ps$^{-1})$  & $\delta G(q)/\gamma(q)$ \\
\hline
0.29711 &   0.9982    &  0.1733    &  -0.009\footnotemark[1]\\
0.70726 &    1.0091   &   0.9222   &   -0.036\footnotemark [1]\\
1.01482 &   1.1292    &  1.7500     &  ~0.173\\
1.09165 &   0.7338     &  1.9753    &  -0.273\\
1.10505 &    0.5122    &   2.0146   &  -0.258\\ 
1.14429 &    1.0576    &    2.1312  &  -0.004\footnotemark[1]\\
1.50523 &    0.7828    &    3.1805  &  -0.075\footnotemark[1] \\
\end{tabular}
\end{ruledtabular}
\label{table II}
\footnotetext[1]{In applying the theory, we set $\delta G/\gamma = 0$ for this $q$.}\
\end{table}

Success of the standard plan in the first peak regime encourages application of the same procedure to the Brillouin peak regime. The challenge here is the Brillouin peak oscillation.
The oscillation itself is present in both $F_{MD}^{d}(q,t)$ and  $F_{vib}^{d}(q,t)$, but it requires an additional damping in the vibrational function to agree with the MD function at $t>t_{c}$. However, in the standard plan, the oscillation is completely removed in $F_{VT}^{d}(q,t)$, via  Eq.~(\ref{eq32}), so the entire oscillation in $F_{MD}^{d}(q,t)$  appears as an error in $F_{VT}^{d}(q,t)$ at $t>t_{c}$. This is not a problem for the standard plan, since the oscillation amplitude is generally less than our theoretical error. 

\begin{figure} [h!]
\includegraphics [width=0.35\textwidth]{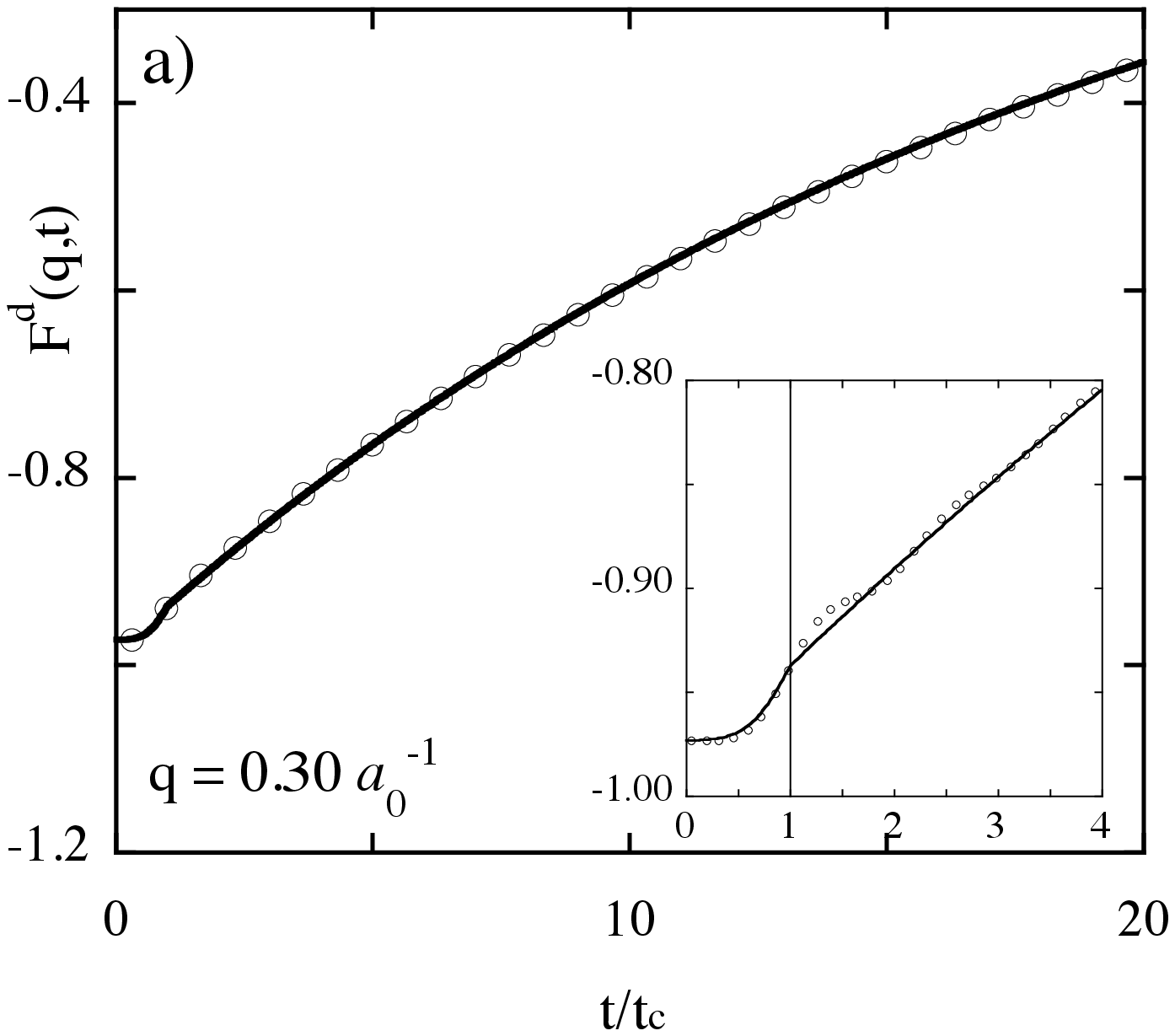}
\includegraphics [width=0.35\textwidth]{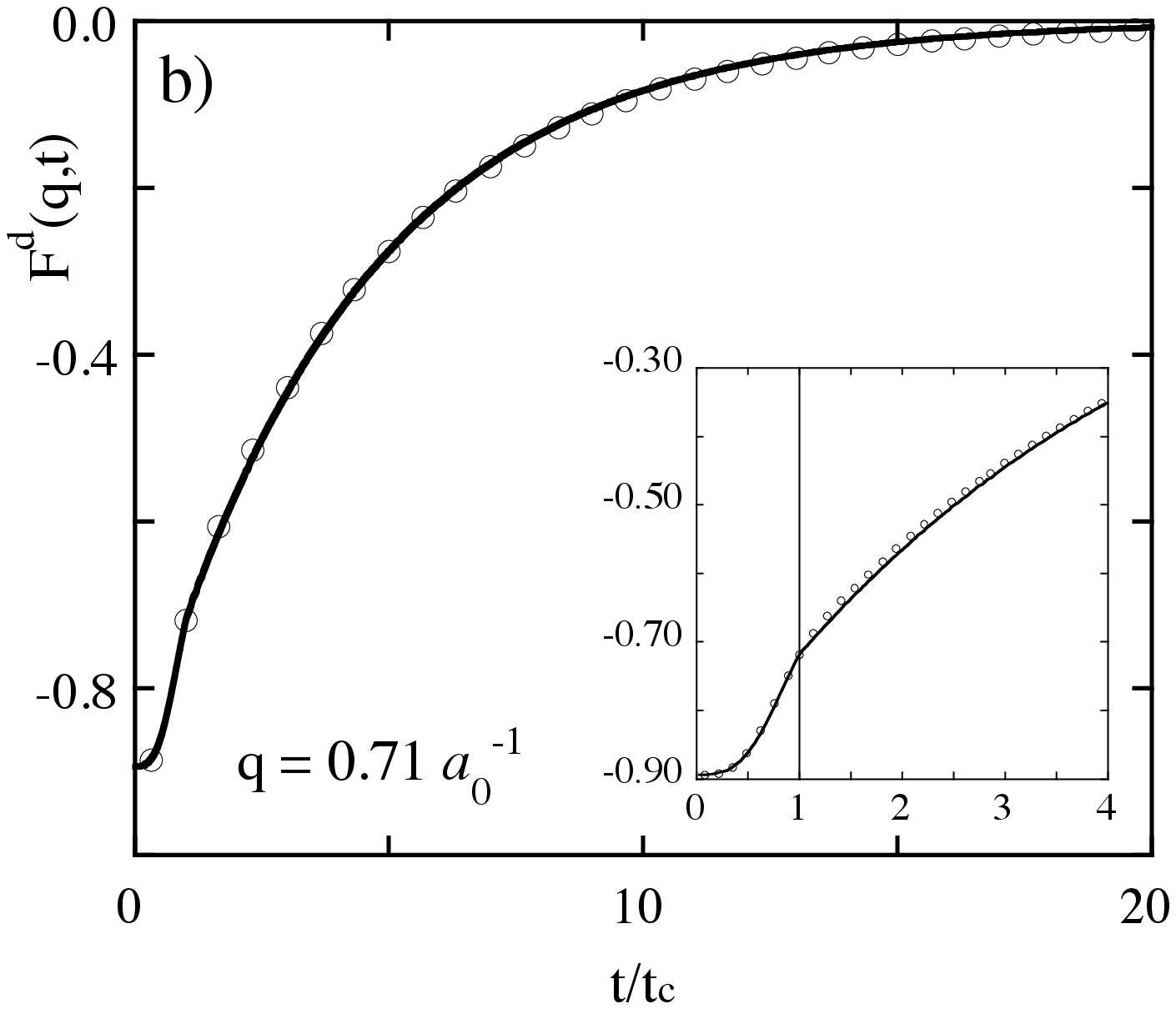}
\caption{For $F^{d}(q,t)$: Comparison of the complete  V-T theory (line) with MD (circles), for two $q$ in the Brillouin peak regime. 
 Inset shows crossover and the Brillouin peak oscillation. }
\label{fig11-12}
\end{figure}

Comparison of theory with MD in the Brillouin peak regime is shown in Fig.~\ref{fig11-12}a and \ref{fig11-12}b for $q=0.30$ and $0.71$, respectively.
In spite of missing the oscillation,  the standard plan remains accurate to $\lesssim 0.01$ in the Brillouin peak regime (Fig.~\ref{fig9}b). It is significant that we can set $\delta G(q)=0$ for both test $q$ in the Brillouin peak regime (see Table~II).

\begin{figure} [h!]
\includegraphics [width=0.35\textwidth]{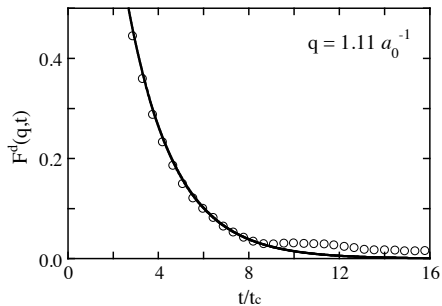}
\caption{For $F^{d}(q,t)$: Long-time data showing the MD tail at $t/t_{c}>8$ (circles), not yet addressed in V-T theory (line). }
\label{fig13}
\end{figure}

The $F_{MD}(q,t)$ data exhibit a long time tail at $q=1.01$ and $1.11$, shown in Fig.~\ref{fig13} at  $q=1.11$ (see also Fig.~\ref{fig9}b). We are investigating the source of this feature. 
Fig.~\ref{fig13} also shows the good agreement of theory with MD as the curves approach zero, before the tail begins.

In the large $q$ regime, at $q$ beyond the first peak, three $q$ were chosen for study, located respectively at the first minimum, second maximum, and second minimum of $F_{MD}^{d}(q,0)$  (see Fig.~\ref{eq1} and Table I). At $q=1.51$, shown in Fig.~\ref{fig14}, it is perhaps surprising that the standard plan still works well. Both motional and structural decorrelation are accurately accounted for, the theoretical slope discontinuity at $t_{c}$ is insignificant, and we are able to set $\delta G(q) =0$ in the structural damping. The deviation $\Delta F^{d}(q,t)$ in Fig.~\ref{fig9} is well below $0.01$. 

\begin{figure} [h!] \label{fig14}
\includegraphics [width=0.35\textwidth]{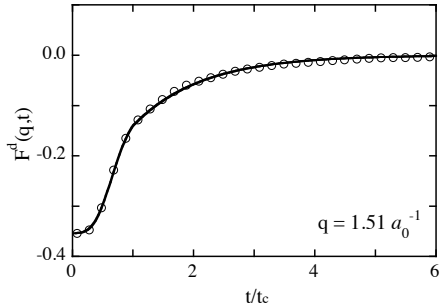}
\caption{For $F^{d}(q,t)$: Comparison of the complete  V-T theory (line) with MD (circles), for  $q=1.51~a_{0}^{-1}$, at the minimum after the first peak. Vertical line is at $t_c$.}
\label{fig14}
\end{figure}

As $q$ increases further, two important changes occur. First, $F_{vib}^{d}(q,\infty)$ goes to zero, as noted following Eq.~(\ref{eq6}), and as seen in the last three entries of Table~I. Second, the decorrelation process becomes faster, with $t_{c}(q)$ falling below the first-transit period of $\nu^{-1}$, again as seen in Table~I. These changes mark the ultimate convergence of the distinct autocorrelation function toward zero as $q$ increases. Both trends are also found in the self autocorrelation function as $q$ increases toward the free particle limit.~\cite{us_Fqtself} As a result, at $q=2.00$ and $2.51$, the total process resembles a single motional decorrelation of an unusual shape. The indication is that $q \gtrsim 2.00$ does not measure structural correlation, and measures motional correlation in a way different from what we have seen at  smaller $q$.

\section{Range of applicability of the standard plan}

\subsection{Exclusion band}There are several $q$-segments where $F_{MD}^{d}(q,0)$ and/or $F_{vib}^{d}(q,0)$ are of small magnitude, say $\lesssim 0.1$, where analysis is difficult for two reasons. First, the values of the MD and vibrational autocorrelation functions are small enough to be compromised by finite-$N$ error, at our present $N=500$. Second, the MD and vibrational functions cross zero at different $q$-values, a phase shift not treated in the standard plan. It makes sense to leave these effects until the dominant part of the theory is developed.

\subsection{First-peak regime}This is where $F_{MD}^{d}(q,0)$ and $F_{vib}^{d}(q,0)$ are positive in the first peak. This regime provided the test case from which the standard plan was developed. Here the $F_{MD}^{d}(q,t)$ curves have uniform character across the $q$-range, and so do the $F_{vib}^{d}(q,t)$ curves. Since distinct and self contributions are both positive, no cancellation complications arise.

\subsection{Brillouin-peak regime}This is where the Brillouin peak oscillation is apparent in $F_{MD}^{d}(q,t)$; in our system it runs from $q=0.13$, the smallest allowed $q$, to somewhat beyond $q=0.80$. Here, $F_{MD}^{d}(q,0)$ and $F_{vib}^{d}(q,0)$ are around $-1$; on this scale the standard plan is accurate, as shown in Fig.~\ref{fig9}, and may be applied.

In this regime, the self and distinct autocorrelation functions nearly cancel for MD data, and also for vibrational data. The Brillouin peak oscillation appears entirely in the distinct function, and contains important physics not addressed in the standard plan, namely the inelastic scattering cross section as function of $q$ and $\omega$. For our system, the Brillouin peak oscillation appears in $F_{vib}^{d}(q,t)$ with the same frequency as it appears in $F_{MD}^{d}(q,t)$, for all $q$ in the regime. Hence $F_{vib}^{d}(q,t)$ already contains the Brillouin peak dispersion curve, \emph{a priori} and to very high accuracy.~\cite{I} What is still required for a complete theory of dynamic response is an extremely accurate decorrelation theory for the Brillouin peak oscillation.

\subsection{Large-$q$ regime} Beyond the first peak, near the first and second minima and the second maximum, where $|F_{MD}^{d}(q,0)|$ is large enough to justify an exploratory investigation, the character of the decorrelation process changes under two influences: The motional decorrelation becomes faster, and $F_{vib}^{d}(q,\infty)$ goes to zero (see Table~I). At $q=1.51$, the standard plan still applies and is still accurate, but a practiced eye will see the beginning of change. At $q=2.00$ and $2.51$, the entire process has the appearance of a single motional decorrelation, an appearance we expect to remain as $q$ increases further, and the MD and vibrational autocorrelation functions converge to zero (see the discussion at the end of Sec.~V). The standard plan does not apply at these $q$.

\section{Summary and Conclusion}

This work is an investigation into the validity of V-T theory for the distinct density autocorrelation function of a monatomic liquid. Our main result is Eqs.~(\ref{eq20}) and (\ref{eq32}), which together accurately model $F^d(q,t)$ over the broad range of $q$ for which the standard decorrelation plan applies. That result depends on a theoretical notion, the transit drift introduced in Sec.~IIIA, which is the basis of our theories of motional and structural decorrelation in $F^d(q,t)$. In this study, the vibrational contribution is parametrized by appeal to a realistic many-body potential for Na, and the various transit parameters are determined from MD simulations using that potential, but we emphasize that this is not necessary: once one understands the physical meaning behind the parameters, their values may be acquired by the method of one's choice. With that in mind, we summarize the physical considerations that led to the model and give the parameters their meaning.

At the outset, the two classes of configurational correlation, motional and structural, are precisely defined by the perfect vibrational system. The motional correlation is measured by the time correlation functions in Eq.~(\ref{eq5}), and the motional decorrelation (due to vibrations and transits) carries these functions from their $t=0$ values to zero. In the process, $F_{vib}^{d}(q,t)$ goes from $F_{vib}^{d}(q,0)$ to $F_{vib}^{d}(q,\infty)$\, according to Eqs.~(\ref{eq3}) and (\ref{eq6}). The structural correlation is measured by the set of equilibrium positions $\{ {\bf R}_{K}\}$, and the Debye-Waller factors $\{W_{K}(\bf{q})\}$, in Eq.~(\ref{eq6}) for $F_{vib}^{d}(q,\infty)$. The ${\bf R}_{K}$ remain fixed during the motional damping, then the ${\bf R}_{K}$ begin to move and proceed to damp $F_{vib}^{d}(q,\infty)$ to zero. This outline of the process is defined entirely in terms of the theoretical function $F_{vib}^{d}(q,t)$, the definition applying to all monatomic liquids.

The three functions $A(q)$, $t_c(q)$, and $G(q)$ that enable calculation of the detailed time dependence are  transit properties.  In physical meaning, $A(q)$ is the transit contribution to the static structure factor. $A(q)$ is supposed to be mainly structural, hence is modeled as a multiple of $F_{vib}^{d}(q,\infty)$. The crossover time $t_{c}(q)$ is the time required to damp the motional correlation to zero, under the simultaneous actions of natural dephasing and transit-induced decorrelation. The physical key to  $t_{c}(q)$ is the condition that long-time motional correlation due to low-lying normal modes is damped out by transits in the liquid. Finally, the structural damping coefficient $G(q)$ expresses the transit random walk, and an \emph{a priori} zeroth-order theory for $G(q)$ is already in place.~\cite{us_Fqtself} All these properties are attributed to the same transit motion in every monatomic liquid.

The transit drift is the massively averaged motion resulting specifically from transits in the liquid state, and present in addition to the normal mode motion. In Sec.~III, in constructing a model equation for the drift-induced motional decorrelation, in the period $0 \leq t \leq t_{c}(q)$, three predictions are made about characteristic properties of the drift. The predictions are expressed in terms of the transit parameters $\nu$ and $\delta R$, previously calibrated independently of any study of $F(q,t)$, with values given in Eq.~(\ref{eq14}). The predictions are listed in Eq.~(\ref{pinco}). 
\begin{enumerate}[a)]
\item  The first prediction is $t_{c}(q) \approx \nu^{-1}$. This is qualitatively correct. Specifically, $\nu^{-1}=0.26$~ps, while from Table~I the average $t_{c}(q)$ for seven $q$ in the standard plan is $0.30\pm 0.10$~ps. 
\item  The second prediction is $s(q,t_{c})\approx \frac{1}{2}\delta R$. This is qualitatively correct. Specifically, $ \frac{1}{2}\delta R=0.88~a_{0}$, while from Table~I the average $s(q,t_{c})$ for seven $q$ in the standard plan is $0.70\pm 0.10~a_{0}$. 
\item The third prediction is $s(q,t) \varpropto t$ for  $0 \leq t \leq t_{c}(q)$. The linear $t$  dependence is used in evaluating the motional decorrelation, Eq.~(\ref{eq22}), and agreement of theory with MD in Fig.~\ref{fig567} verifies the $t$ dependence within the observed errors. 
\end{enumerate}
The level of agreement is excellent for nonequilibrium data; further, the systematic character of the V-T theory errors shown in Fig.~\ref{fig9}a indicates that the motional decorrelation theory of Eqs.~(\ref{eq19}) and (\ref{eq20}) can still be improved.

The transit drift is a steady state motion, hence is always present in the liquid. However, in order to address the structural decorrelation at $t > t_{c}$, it is necessary to resolve the drift into its effective motion of the equilibrium positions. This resolution yields the transit random walk. Indeed, the time required after $t=0$ for the random walk to come into effect provides the physical explanation for the delay to $t_{c}(q)$ of the onset of structural decorrelation, as it appears in Eq.~(\ref{eq32}).

Simple exponential decay, beginning after a delay period, is a hallmark of MD data for self and distinct density autocorrelation functions. In self decorrelation, the theoretical damping coefficient $\gamma (q)$ is in excellent agreement with MD data for all $q$. The result constitutes a unification within V-T theory: One cannot calculate the density autocorrelation damping coefficient $\gamma(q)$ from the self diffusion coefficient $D$,~\cite{HMcD_3rded, BZ1994} but one \emph{can} calculate both $\gamma (q)$ and $D$ from the same random walk of the equilibrium positions.

Finally, we shall make a highly approximate model for the anistropic  damping coefficient $\delta G(q)$. Consider a representative pair of atoms $K,L$ which are nearest neighbors at $t=0$. Because the atomic dynamics enforces a minimum distance between atoms, the first-transit drift of atom $K$ will not be isotropically distributed about ${\bf R}_{K}(0)$, but will move preferentially away from ${\bf R}_{L}(0)$. The transit random walk will not develop about ${\bf R}_{K}(0)$, but about ${\bf R}_{K}(0) + \delta {\bf{s}}_{K}$, where $\delta {\bf{s}}_{K}$ is a small displacement in the direction of ${\bf R}_{KL}(0)$. The picture can be developed from Eq.~(\ref{eq17}). The resulting $\delta G(q)$ will be positive for $q$ on the first-peak leading edge, and negative at the first-peak tip. This is what we see in Sec.~IV. Moreover, since the effect appears only for nearest neighbors, $\delta G/\gamma$ will be significant only for $q$ in the first peak regime. Even for $q$ in the first peak, $\delta G/\gamma$ is small because the correlation is weak after the first transit. All these details of $\delta G/\gamma$ are confirmed by the last column of Table~II. The theoretical challenge now is a more quantitative understanding of $\delta G/\gamma$.

Our expressions for $F^d(q, t)$ rely on the vibrational parameters $\{{\bf{R}}_K, \omega_\lambda, {\bf{w}}_{K\lambda}\}$ and three transit functions $A(q), t_{c}(q)$, and $G(q)$. In this study, the vibrational parameters were determined from a single random valley of the Na pseudopotential, and the transit functions were determined from combined MD and vibrational data. This is done in order to show the extreme accuracy of the calibrated theory; we believe this high-accuracy capability gives credence to the formulation. On the other hand, the vibrational parameters can be determined from first principles: examine a random valley on a potential surface calculated from DFT.  The remaining functions are determined by transit dynamics and can in principle be calculated from a transit model. (The current transit model already provides reasonably accurate estimates $\nu ^{-1}$ for $t_{c}(q)$, and $\gamma (q)$ for $G(q)$.) An improved transit model, in addition to providing better estimates for $t_{c}(q)$ and $G(q)$, would ideally (a) describe the transit decorrelation processes that dominate for $q$ outside the standard plan and (b) explain the origin of the initial transit-based correlations, given by $A(q)$.


\begin{thebibliography}{51}
\bibitem[{fqt()}]{fqtd}
\bibinfo{howpublished}{``Distinct autocorrelation function" is an abbreviation
  for ``distinct contribution to the density autocorrelation function"}.
\end{thebibliography}

\end{document}